\documentclass[%
 aip,
 amsmath,amssymb,
 reprint,%
]{revtex4-1}

\usepackage{parskip}
\usepackage{units}
\usepackage{graphicx}
\usepackage{dcolumn}
\usepackage{bm}
\usepackage{moresize}
\usepackage{amsmath}
\usepackage[utf8]{inputenc}
\usepackage[T1]{fontenc}
\usepackage{mathptmx}
\usepackage{etoolbox}
\usepackage[colorlinks]{hyperref}
\hypersetup{%
	plainpages=true,
	breaklinks=true,
	hypertexnames=false,
	pageanchor=true,
	colorlinks=true,
	linkcolor={blue},
	citecolor={blue},
	urlcolor={blue},
	anchorcolor={black}}

\newcommand{\ket}[1]{|#1\rangle}

\usepackage{mleftright} 

\newcommand{\figref}[1]{\mbox{Fig.~\ref{#1}}}

\renewcommand{\eqref}[1]{\mbox{Eq.~(\ref{#1})}}

\newcommand{\figpanel}[2]{Fig.~\hyperref[#1]{\ref*{#1}(#2)}}
\newcommand{\figpanels}[3]{Fig.~\hyperref[#1]{\ref*{#1}(#2)-(#3)}}
\newcommand{\figpanelNoPrefix}[2]{\hyperref[#1]{\ref*{#1}(#2)}}

\makeatletter
\def\@email#1#2{%
 \endgroup
 \patchcmd{\titleblock@produce}
  {\frontmatter@RRAPformat}
  {\frontmatter@RRAPformat{\produce@RRAP{*#1\href{mailto:#2}{#2}}}\frontmatter@RRAPformat}
  {}{}
}%
\makeatother
\begin{document}

\preprint{AIP/123-QED}


\title{Tunable frequency conversion and comb generation \\ with a superconducting artificial atom}

\author{Fahad Aziz}
\thanks{These authors contributed equally to this work.}
\affiliation{Department of Physics, National Tsing Hua University, Hsinchu 30013, Taiwan}
\affiliation{Department of Physics, City University of Hong Kong, Kowloon, Hong Kong SAR 999077, China}

\author{Zhengqi~Niu}
\thanks{These authors contributed equally to this work.}
\affiliation{State Key Laboratory of Materials for Integrated Circuits, Shanghai Institute of Microsystem and Information Technology (SIMIT), Chinese Academy of Science, Shanghai 200050, China}

\author{Tzu-Yen~Hsieh}
\thanks{These authors contributed equally to this work.}
\affiliation{Department of Physics and CQSE, National Taiwan University, Taipei 10617, Taiwan}%

\author{Kuan Ting~Lin}
\thanks{These authors contributed equally to this work.}
\affiliation{Trapped-Ion Quantum Computing Laboratory, Hon Hai Research Institute, Taipei 11492, Taiwan}

\author{Yu-Huan Huang}
\affiliation{Department of Physics, National Tsing Hua University, Hsinchu 30013, Taiwan}

\author{Yen-Hsiang Lin}
\affiliation{Department of Physics, National Tsing Hua University, Hsinchu 30013, Taiwan}
\affiliation{Center for Quantum Technology, National Tsing Hua University, Hsinchu 30013, Taiwan}

\author{Ching-Yeh Chen}
\affiliation{Department of Physics, National Tsing Hua University, Hsinchu 30013, Taiwan}

\author{Yu-Ting Cheng}
\affiliation{Department of Physics, City University of Hong Kong, Kowloon, Hong Kong SAR 999077, China}

\author{Kai-Min Hsieh}
\affiliation{Department of Physics, City University of Hong Kong, Kowloon, Hong Kong SAR 999077, China}

\author{Jeng-Chung Chen}
\affiliation{Department of Physics, National Tsing Hua University, Hsinchu 30013, Taiwan}
\affiliation{Center for Quantum Technology, National Tsing Hua University, Hsinchu 30013, Taiwan}

\author{Anton Frisk Kockum}
\affiliation{Department of Microtechnology and Nanoscience, Chalmers University of Technology,  412 96 Gothenburg, Sweden}

\author{Guin-Dar Lin}
\affiliation{Department of Physics and CQSE, National Taiwan University, Taipei 10617, Taiwan}
\affiliation{Trapped-Ion Quantum Computing Laboratory, Hon Hai Research Institute, Taipei 11492, Taiwan}
\affiliation{Physics Division, National Center for Theoretical Sciences, Taipei 10617, Taiwan}

\author{Zhi-Rong~Lin}
\thanks{Authors to whom correspondence should be addressed: \\
zrlin@mail.sim.ac.cn; \\
eighchin@ccu.edu.tw; \\
and iochoi@cityu.edu.hk}
\affiliation{State Key Laboratory of Materials for Integrated Circuits, Shanghai Institute of Microsystem and Information Technology (SIMIT), Chinese Academy of Science, Shanghai 200050, China}

\author{Ping-Yi Wen}
\thanks{Authors to whom correspondence should be addressed: \\
zrlin@mail.sim.ac.cn; \\
eighchin@ccu.edu.tw; \\
and iochoi@cityu.edu.hk}
\affiliation{Department of Physics, National Chung Cheng University, Chiayi 621301, Taiwan}

\author{Io-Chun Hoi}
\thanks{Authors to whom correspondence should be addressed: \\
zrlin@mail.sim.ac.cn; \\
eighchin@ccu.edu.tw; \\
and iochoi@cityu.edu.hk}
\affiliation{Department of Physics, National Tsing Hua University, Hsinchu 30013, Taiwan}
\affiliation{Department of Physics, City University of Hong Kong, Kowloon, Hong Kong SAR 999077, China}

\date{\today}


\begin{abstract}

We investigate the power spectral density emitted by a superconducting artificial atom coupled to the end of a semi-infinite transmission line and driven by two continuous radio-frequency fields. In this setup, we observe the generation of multiple frequency peaks and the formation of frequency combs with equal detuning between those peaks. The frequency peaks originate from wave mixing of the drive fields, mediated by the artificial atom, highlighting the potential of this system as both a frequency converter and a frequency-comb generator. We demonstrate precise control and tunability in generating these frequency features, aligning well with theoretical predictions, across a relatively wide frequency range (tens of MHz, exceeding the linewidth of the artificial atom). The extensive and simple tunability of this frequency converter and comb generator, combined with its small physical footprint, makes it promising for quantum optics on chips and other applications in quantum technology.

\end{abstract}



\maketitle 


\textit{Introduction.} In the past two decades, superconducting circuits~\cite{blais2021circuit} with artificial atoms coupled to open waveguides, one platform for waveguide quantum electrodynamics (wQED)~\cite{Roy2017, gu2017microwave}, has provided a new level of understanding of quantum light-matter interaction~\cite{sheremet2021waveguide}. These systems not only advance fundamental quantum mechanics~\cite{Roy2017, gu2017microwave, sheremet2021waveguide, Astafiev2010, PhysRevLett.104.183603, hoi2013giant, vanLoo2013, hoi2015probing, PhysRevLett.123.233602, Mirhosseini2019, kannan2020waveguide, wen2020landau, lu2021quantum, Kockum2021, wiegand2021ultimate, Wei-Ju, lu2022steady, cheng2024group, aziz2025} but also enable practical quantum technologies~\cite{PhysRevLett.107.073601, kono2024microwave, Chen2025, Almanakly2025, huang2025tunable}. Among building blocks for quantum technology, frequency conversion~\cite{kumar1990quantum, Han21, Jianming} and frequency-comb generation~\cite{yamazaki2023linear, chang2022integrated, jolin2023multipartite, rivera2024control} have gained significant interest across diverse areas, with applications ranging from quantum networks to precision measurements~\cite{ludlow2015optical, cundiff2003colloquium}. For implementing these building blocks through wave mixing, wQED provides an integrated on-chip platform. Compared to setups in traditional optics, the wQED approach reduces complexity and size of components, enhances scalability, and can reach lower noise and losses.

Recent advances in wave mixing with superconducting qubits have yielded encouraging results, including quantum wave mixing (QWM)~\cite{dmitriev2017quantum}, coherent wave mixing~\cite{honigl2018mixing}, photon statistics of coherent states~\cite{dmitriev2019probing}, and amplification without population inversion~\cite{wen2018reflective}. Prior studies analyzing power spectral density (PSD) in these systems have typically employed pulsed sequences~\cite{dmitriev2017quantum} or continuous waves~\cite{dmitriev2019probing} with equal spectrum width ($\Delta\omega$) within the qubit's linewidth ($\Gamma$), such that $\Gamma/2\pi \gg \Delta\omega/2\pi  \sim \unit[]{kHz}$. This constraint limits the performance of both frequency conversion and frequency-comb generation. In Ref.~\cite{honigl2018mixing}, all frequencies involved are centered around the transition frequencies of a three-level atom, which constrains the generated frequencies. The setup of Ref.~\cite{wen2018reflective} differs in that it measures the reflection coefficient, which primarily captures elastic scattering; the full information on spectral lines is missing.

\begin{figure}
\includegraphics[width=0.99\linewidth]{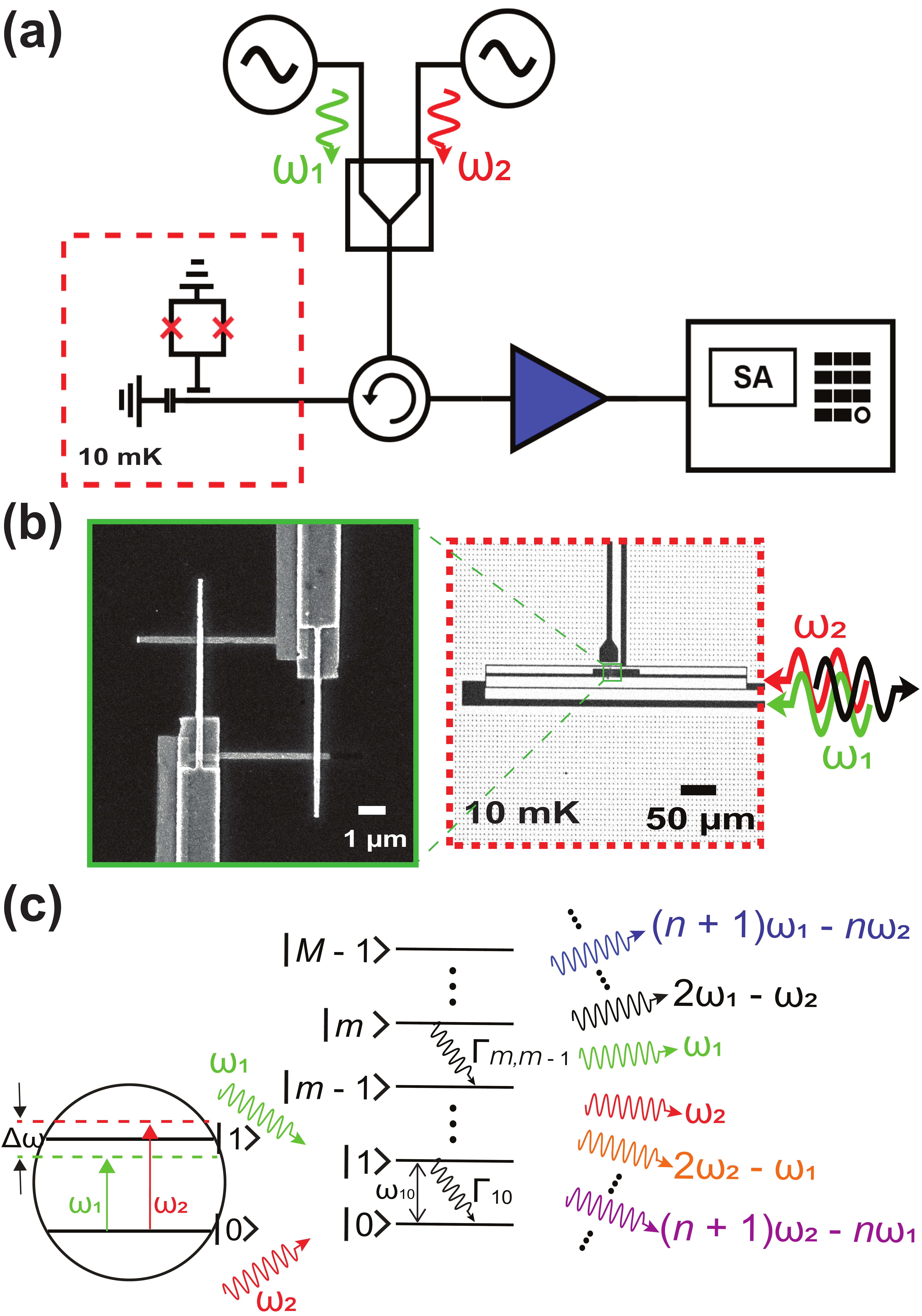}
\caption{Experimental scheme, sample, and theoretical model.
(a) Simplified schematic of the experimental setup, featuring two input fields, with carrier frequencies $\omega_1$ and $\omega_2$, combined via an RF combiner in adding mode at room temperature. The sample (red dashed box) is placed at the bottom of a dilution refrigerator at \unit[10]{mK}. The input and output fields are separated by a cryogenic circulator, with the output field amplified and measured by a spectrum analyzer (SA; see the Supplementary Material~\cite{Supp} Section S1 for details). 
(b) Device used in this experiment. A single artificial atom (transmon), a superconducting circuit, is capacitively coupled to the end of a semi-infinite transmission line~\cite{wen2018reflective, PhysRevA.109.023705}. The vertical line is a local flux line that is not utilized in this work. We employ only a single input-output port, as indicated by the incident (red and green) and reflected (black) fields. The small green square marks the position of the superconducting quantum interference device (SQUID); the inset is a scanning electron microscope (SEM) image of the SQUID, which has two Josephson junctions.
(c) An $M$-level transmon, with $M = 5$, serving as a nonlinear medium, is pumped by the two continuous RF fields with carrier frequencies $\omega_1$ and $\omega_2$, respectively, with frequency difference $\Delta \omega = \omega_2 - \omega_1$. The detailed theoretical model is presented in Section S4 of the Supplementary Material~\cite{Supp}.
\label{Fig1}}
\end{figure}

Here, we drive a superconducting artificial atom coupled to the end of a semi-infinite transmission line, which effectively collects all the emission in one direction~\cite{aziz2025, wiegand2021ultimate}, with two continuous radio-frequency (RF) fields; see the experimental setup in \figpanel{Fig1}{a,b}. We analyze the PSD of the emission over a broad frequency range near $\omega_{10}$, the $\ket{0} \leftrightarrow \ket{1}$ transition frequency of the atom, where $\ket{0}$ is the ground state and $\ket{1}$ is the first excited state. We utilize the nonlinear interactions between the drive fields and the transmon to demonstrate a tunable frequency converter all the way from $\unit[1]{MHz}$ to $\unit[50]{MHz}$ and a frequency-comb generator with $\Delta \omega / 2\pi \sim \unit[1]{MHz}$--$\unit[60]{MHz}$. By controlling the RF power and frequency, we achieve precise tuning of new frequency peaks and the comb generation, aligning with theoretical predictions [see \figpanel{Fig1}{c} for an illustration of the theoretical model of the operation of the system] and expanding the capabilities of such systems. 

\textit{Device characterization}. First, we characterize the artificial atom to determine its fundamental parameters. This characterization is performed using single-tone spectroscopy with a vector network analyzer (VNA) to measure the magnitude $|r|$ or the amplitude reflection coefficient (see Section S2 of the Supplementary Material~\cite{Supp} for details). The key parameters we extract for the $\ket{0} \leftrightarrow \ket{1}$ transition are the transition frequency $\omega_{10} / 2\pi = \unit[4.82]{GHz}$, the relaxation rate $\Gamma_{10} / 2\pi = \unit[44.2]{MHz}$, the decoherence rate $\gamma_{10} / 2\pi = \unit[22.5]{MHz}$, and the pure dephasing rate $\Gamma_{\phi} / 2\pi = \unit[0.37]{MHz}$. These parameters are crucial for understanding the behavior of the artificial atom and comparing the theoretical model to the results of the experiments. We also performed two-tone spectroscopy to measure the reflection coefficient $|r|$ as a function of drive power, allowing us to observe the Mollow triplet (see Section S3 of the Supplementary Material~\cite{Supp} for details) and to extract the transmon anharmonicity $(\omega_{10} - \omega_{21}) / 2\pi \cong E_C / h \cong \unit[320]{MHz}$, where $E_C$ is the transmon charging energy.

\textit{Frequency up- and down-conversion}.
We now switch to the experimental scheme shown in \figpanel{Fig1}{a}, wherein we input two continuous RF fields with carrier frequencies $\omega_1$ and $\omega_2$ into the system. The fields are combined using an RF combiner and subsequently attenuated at each stage of the dilution refrigerator before being directed through the circulator to interact with the artificial atom. The reflected output, which is the field emitted after interacting with the artificial atom, is captured and analyzed using a spectrum analyzer.

To measure the normalized power spectral density PSD$_n$, we first record the output trace with both RF fields ($\omega_1$ and $\omega_2$) turned on and then with the fields turned off. We subtract the trace with the RF fields off from the trace with the RF fields on to obtain PSD$_n$. We employ this method consistently throughout the manuscript to analyze new frequency features arising from the mixing of the carrier frequencies $\omega_1$ and $\omega_2$, as depicted in \figpanel{Fig1}{c}. The transmon's energy-level differences are theoretically approximated as $\omega_{m, m-1} = \omega_m - \omega_{m-1} \approx \omega_{m-1, m-2} + (1 - m) E_C / \hbar$ for $m = 1, 2, \ldots, 5$. This energy nonlinearity enables the generation of output frequencies that are combinations of the input fields, including the fundamental frequencies ($\omega_1$, $\omega_2$), the first harmonic frequencies ($2 \omega_1 - \omega_2$, $2 \omega_2 - \omega_1$), and higher-order mixed frequencies [$(n+1) \omega_1 - n \omega_2$, $(n+1) \omega_2 - n \omega_1$]. The relaxation rate from $\ket{m}$ to $\ket{m-1}$ is given by $\Gamma_{m, m-1} = m \Gamma_{10}$. See Section S4 of the Supplementary Material~\cite{Supp} for the detailed theoretical model.

\begin{figure*}
\includegraphics[width=\linewidth]{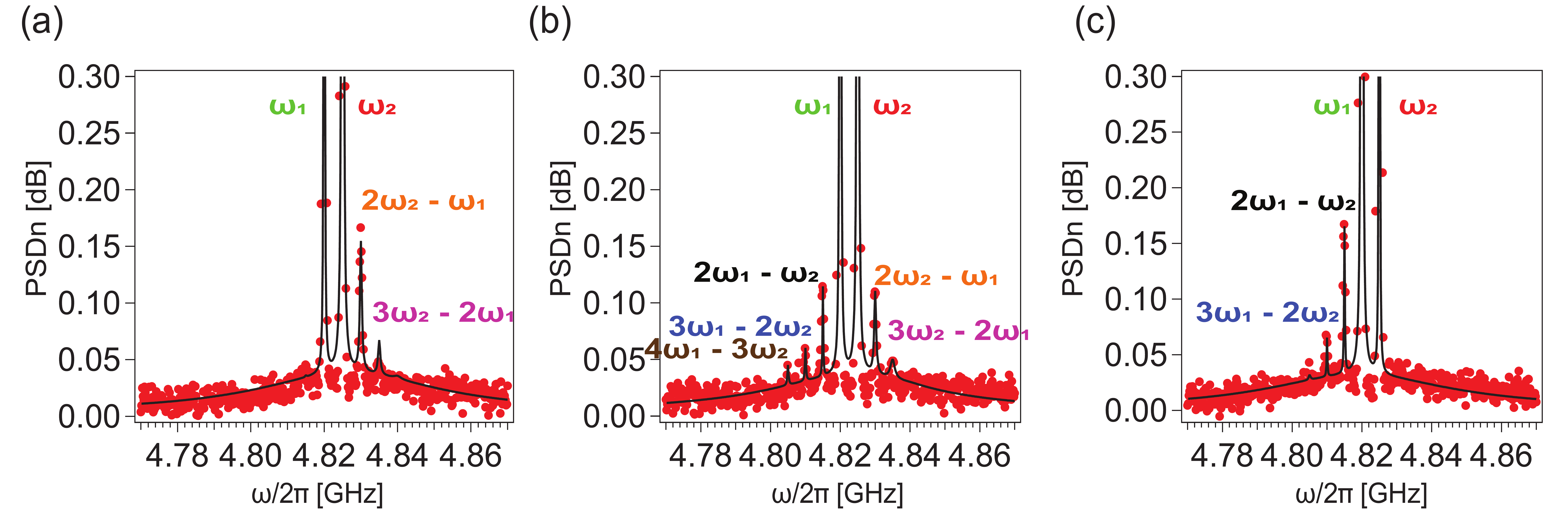}
\caption{\label{Fig2} Frequency up- and down-conversion. Two continuous fields at frequencies $\omega_1$ and $\omega_2$ are applied with a fixed power of $P_2 = \unit[-125]{dBm}$, and the normalized power spectral density is measured under three distinct input powers $P_1$:
(a) $\unit[-129]{dBm}$, 
(b) $\unit[-125]{dBm}$, and 
(c) $\unit[-121]{dBm}$. 
Three notable features emerge. In (a), where $P_1 < P_2$, a new spectral peak appears at $2 \omega_2 - \omega_1$. In (b), where $P_1 = P_2$, two symmetrical peaks form on both sides of the carrier frequencies $\omega_1$ and $\omega_2$, each at equal detuning. In (c), where $P_1 > P_2$, a new peak arises at $2 \omega_1 - \omega_2$. Red dots represent experimental data, and solid black curves depict theoretical predictions, showing good agreement.
}%
\end{figure*}

Figure~\ref{Fig2} presents our results for frequency conversion. In \figpanel{Fig2}{a}, $\omega_1$ is set to $\omega_{10}$ ($\omega_1 / 2\pi = \unit[4.82]{GHz}$) with an input power of $P_1 = \unit[-129]{dBm}$. Simultaneously, $\omega_2$ is set to a frequency \unit[5]{MHz} blue-detuned (i.e., $\omega_2 / 2\pi = \unit[4.825]{GHz}$) with a higher input power of $P_2 = \unit[-125]{dBm}$. Under these conditions, we observe the emergence of new peaks in the PSD$_n$ at $2 \omega_2 - \omega_1$ and $3 \omega_2 - 2 \omega_1$. These peaks arise from the nonlinear mixing of the carrier frequencies $\omega_1$ and $\omega_2$.

In \figref{Fig2}{b}, we apply both fields with equal input power, $P_1 = P_2 = \unit[-125]{dBm}$. This configuration leads to the observation of prominent new peaks on both sides of the carrier frequencies $\omega_1$ and $\omega_2$. These new peaks are generated at $3 \omega_2 - 2 \omega_1$, $2 \omega_2 - \omega_1$, $2 \omega_1 - \omega_2$, $3 \omega_1 - 2 \omega_2$ and $4 \omega_1 - 3 \omega_2$. These peaks exhibit an equal detuning (peak-to-peak frequency difference) of $\unit[5]{MHz}$. The symmetrical appearance of these peaks (except $4 \omega_1 - 3 \omega_2$) reflects the nonlinear interactions between the RF fields and the artificial atom, providing insight into the frequency-mixing processes, as depicted in \figpanel{Fig1}{c}.

Figure~\figpanelNoPrefix{Fig2}{c} shows the results when we increase the applied power to $P_1 = \unit[-121]{dBm}$, while maintaining $P_2 = \unit[-125]{dBm}$. This scenario mirrors the configuration shown in \figpanel{Fig2}{a}. In this case, the applied power $P_1$ is larger than the applied power of $P_2$, whereas in \figpanel{Fig2}{a}, $P_1 < P_2$. Consequently, we observe new peaks at $2 \omega_1 - \omega_2$ and $3 \omega_1 - 2 \omega_2$. The dots represent experimental data, while the solid curves are theoretical predictions, showing good agreement between the two. The complete experimental and theoretical data are detailed in Section S5, Fig.~S4(a) and (b), of the Supplementary Material~\cite{Supp}. The emergence of these new frequency peaks highlights how their appearance can be controlled by varying the powers $P_1$ and $P_2$. This simple control through modification of the powers of the input signals has potential applications for switching between frequency up-conversion and down-conversion in practical scenarios.

\begin{figure*}
\includegraphics[width=\linewidth]{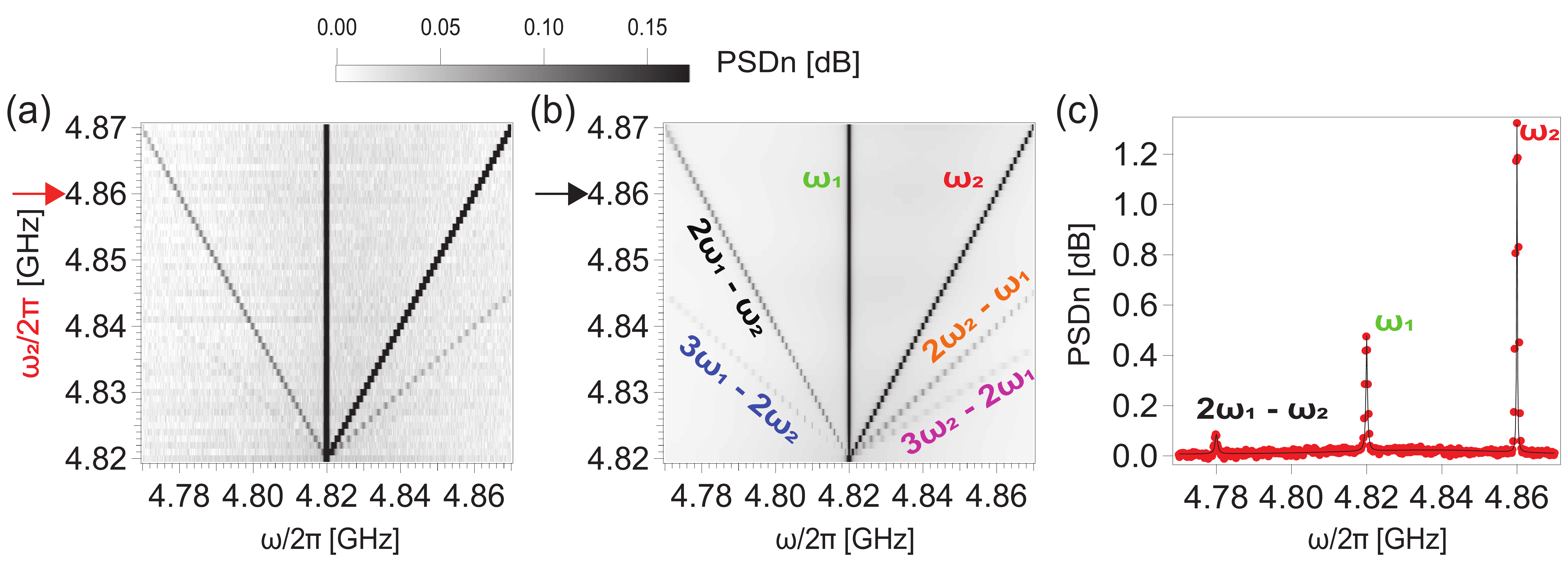}
\caption{\label{Fig3} Tunable frequency conversion. 
(a) Measured PSD$_n$ as a function of spectrum analyzer frequency (x axis) and input frequency $\omega_2$ (y axis). The carrier frequency $\omega_1$ is fixed at $\omega_{10}$ with an input power of  $P_1 = \unit[-130]{dBm}$, while the carrier frequency $\omega_2$ is swept at a constant input power $P_2 = \unit[-130]{dBm}$.
(b) Theoretical prediction for (a). The spectrum reveals four new frequency peaks at $3 \omega_2 - 2 \omega_1$, $2 \omega_2 - \omega_1$, $2 \omega_1 - \omega_2$, and $3 \omega_1 - 2 \omega_2$. 
(c) Line cut at $\omega_2 / 2\pi = \unit[4.86]{GHz}$ [red arrow in (a), black arrow in (b)], displaying a new peak at $2 \omega_1 - \omega_2$. The other peaks are not seen here due to the limited frequency span. The conversion frequency is tunable by selecting an appropriate $\omega_2$ frequency as needed. Red dots represent experimental data and the solid black curve is the theoretical prediction; the two agree well.
}%
\end{figure*}

\textit{Tunable frequency conversion}.
To further explore the tunability of the generated frequency peaks, we fix the power levels of both RF sources at $P_1 = P_2 = \unit[-130]{dBm}$ and sweep the frequency $\omega_2 / 2\pi$ from \unit[4.82]{GHz} to \unit[4.87]{GHz}, while maintaining $\omega_1 / 2 \pi = \omega_{10} / 2 \pi = \unit[4.82]{GHz}$. The results of this experiment are presented in \figpanel{Fig3}{a}, where we clearly observe two new lines at $2 \omega_1 - \omega_2$ and $2 \omega_2 - \omega_1$. These results illustrate the ability to generate and tune new frequency peaks by adjusting the carrier frequency of $\omega_2$ as required while maintaining the $\omega_1$ frequency. Figure~\figpanelNoPrefix{Fig3}{b} presents the corresponding theoretical results, showing four newly generated components in the spectrum. In the experimental data, the higher-order peaks are less prominent than in the theory simulations, but the agreement between theory and experiment is good.

In \figpanel{Fig3}{c}, we show a line cut at $\omega_2 / 2\pi = \unit[4.86]{GHz}$, which is marked by a red arrow in \figpanel{Fig3}{a} and a black arrow in \figpanel{Fig3}{b}. This line cut reveals a generated peak at $2 \omega_1 - \omega_2$, whereas $\omega_1 = \omega_{10}$ and $\omega_2$ is about $2 \gamma_{10}$ away (i.e., $\Delta \omega = 2 \gamma_{10}$), with equal detuning. Note that the peak at $\omega_1$ is much weaker than that at $\omega_2$ because of resonance scattering. The capability to control and tune the frequency conversion beyond the linewidth of the qubit transition is a significant finding, demonstrating the versatility of our setup for applications in frequency generation and manipulation. Additionally, the system's compact footprint allows integration with other quantum devices on a single chip.

\begin{figure*}
\centering
\includegraphics[width=\linewidth]{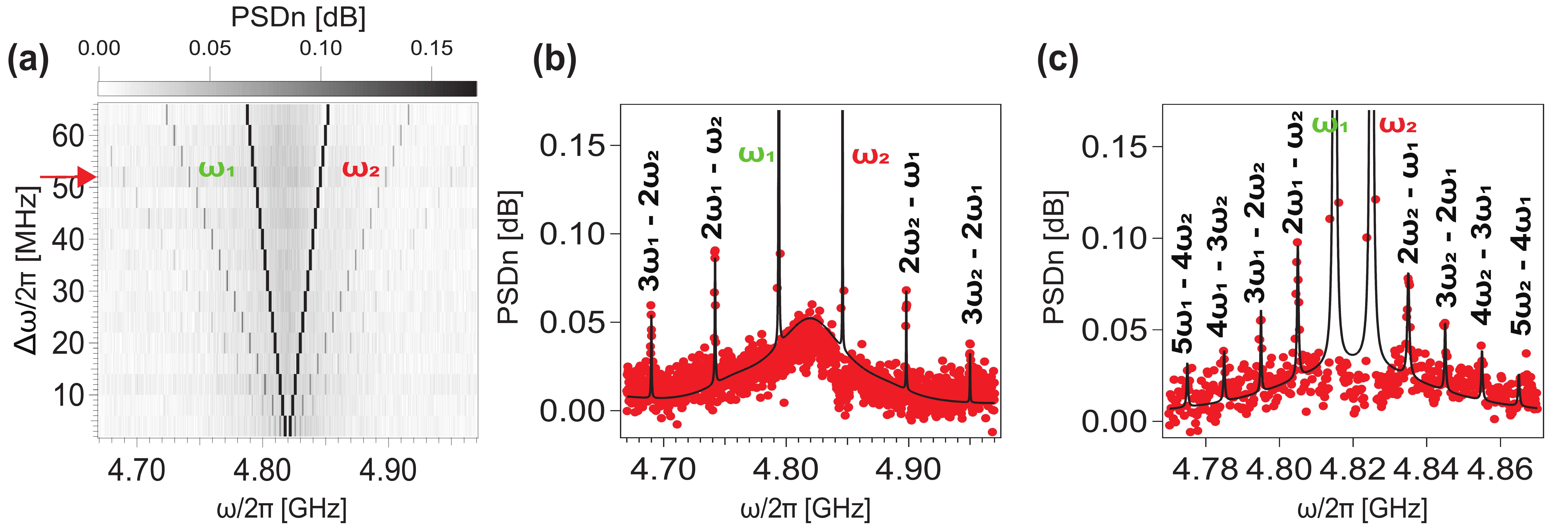}
\caption{\label{Fig4} Generation of frequency combs.
(a) Measured PSD$_n$ as a function of spectrum analyzer frequency (x axis) and detuning $\Delta \omega = \omega_2 - \omega_1$ (y axis). The detuning varies at a fixed input power of $P_1 = P_2 = \unit[-123]{dBm}$. The artificial atom is biased at its resonant frequency, $\omega_{10} / 2\pi = \unit[4.82]{MHz}$, halfway between $\omega_1$ and $\omega_2$. Four new frequencies are generated symmetrically around the carrier frequencies $\omega_1$ and $\omega_2$. 
(b) Line cut from at $\unit[52]{MHz}$, indicated by the red arrow in (a), where all four new peaks are observed.
(c) With input powers increased to $P_1 = P_2 = \unit[-120]{dBm}$ and detuning set to $\unit[10]{MHz}$, eight new peaks emerge at equidistant intervals, though some exhibit small amplitudes. Each peak is labeled by its mixing frequency. This demonstrates frequency-comb generation through appropriate detuning and input power selection. Red dots are experimental data and the solid black curves are theoretical predictions; the two agree well.
}%
\end{figure*}

\textit{Frequency comb}.
Next, we explore the generation of frequency combs. In \figpanel{Fig4}{a}, we sweep the detuning $\Delta \omega = \omega_2 - \omega_1$ on the y axis. The artificial atom is still biased at $\omega_{10} / 2\pi = \unit[4.82]{GHz}$ and both input frequencies are set close to it, such that $\omega_{10} \approx (\omega_1 + \omega_2) / 2$. The input powers $P_1$ and $P_2$ are both set to $\unit[-123]{dBm}$. The x axis represents the frequency range of the spectrum analyzer ($\omega / 2\pi$). We observe the appearance of four new frequency peaks, symmetrically distributed around the input frequencies. As we sweep the detuning, we also observe the broad emission center at $\omega_{10}$ between the carrier frequencies $\omega_1$ and $\omega_2$. This signal comes from the inelastic (incoherent) scattering of the atom (see Section S4~C of the Supplementary Material~\cite{Supp} for details). Figure~\figpanelNoPrefix{Fig4}{b} is a line cut from \figpanel{Fig4}{a} at a detuning of $\unit[52]{MHz}$, marked by a red arrow in \figpanel{Fig4}{a} and by a black arrow in Fig.~S5 in Section S6 of the Supplementary Material~\cite{Supp}. The red dots are experimental data, while the black solid curve is the theoretical prediction, showing good agreement, with a total of six peaks at equidistant intervals of $\unit[52]{MHz}$ detuning. This observation confirms the generation of a frequency comb with distinct frequency peaks.

As the detuning decreases, more frequency peaks are generated. In \figpanel{Fig4}{c}, we set the input powers of both RF sources to $P_1 = P_2 = \unit[-120]{dBm}$, with carrier frequencies $\omega_1 / 2\pi = \unit[4.815]{GHz}$ and $\omega_2 / 2\pi = \unit[4.825]{GHz}$ (see Section S6~B of the Supplementary Material~\cite{Supp} for details). As a result, we observe a total of ten frequency peaks with a detuning of $\unit[10]{MHz}$. Strong input power levels and small detuning lead to a more pronounced generation of frequency combs, demonstrating that the number of frequency peaks can be controlled by adjusting the RF power and detuning. This capability to generate and control frequency combs highlights the potential of our system for applications in precision measurements and frequency synthesis.

Our system also offers several advantages over conventional approaches. In most studies~\cite{cheng2024frequency, gao2024dynamic, szabados2020frequency, wang2024optomechanical, Mrozowski, udem2002optical}, a physical cavity is essential, and the frequency spacing of the comb is constrained by the cavity's free spectral range. In contrast, our frequency-comb generator and frequency converter are cavity-free, providing enhanced tunability and design flexibility. Furthermore, while traditional frequency combs rely on active mode-locking techniques to ensure phase coherence, our system utilizes wave mixing, where the in-phase condition of the four interacting fields is intrinsically satisfied due to momentum conservation. This leads to self-sustained mode locking without the need for external synchronization.

\textit{Conclusion}.
We have demonstrated the effective use of a superconducting artificial atom for tunable frequency conversion and frequency-comb generation across a wide range of frequencies (exceeding the atom linewidth). By analyzing the mixing between two continuous RF fields and the artificial atom, we demonstrate important features, including the generation of new frequency peaks and the production of frequency combs with varying numbers of peaks. We show that we can control all these features by simply varying the input powers and carrier frequencies of the two drive fields. These results align well with our theoretical predictions, confirming the potential of our system for advanced applications in quantum information processing and precise frequency manipulation. Our findings pave the way for further exploration of on-chip microwave frequency-conversion techniques and the development of applications in quantum technology.


\begin{acknowledgments}

I.-C.H.~acknowledges financial support from City University of Hong Kong through the start-up project 9610569, from the Research Grants Council of Hong Kong (Grant number 11312322), and from the Guangdong Provincial Quantum Science Strategic Initiative (Grant No. GDZX2203001, GDZX2303005, GDZX2403001).
A.F.K.~acknowledges support from the Swedish Foundation for Strategic Research (grant numbers FFL21-0279 and FUS21-0063), the Horizon Europe programme HORIZON-CL4-2022-QUANTUM-01-SGA via the project 101113946 OpenSuperQPlus100, and from the Knut and Alice Wallenberg Foundation through the Wallenberg Centre for Quantum Technology (WACQT).

\end{acknowledgments}


\section*{AUTHOR DECLARATIONS}

\subsection*{Conflict of Interest}

The authors have no conflicts of interest to disclose.

\subsection*{Author Contributions}

Fahad Aziz: Experiment (Lead); Writing -- original draft (Lead). Zhengqi Niu: Sample fabrication (Lead). Tzu-Yen Hsieh: Theory (equal). Kuan Ting Lin: Theory (equal). Yu-Huan Huang: Experiment (supporting). Yen-Hsiang Lin: Supervision (supporting); Writing -- review (supporting). Ching-Yeh Chen: Experiment (supporting). Yu-Ting Cheng: Experiment (supporting). Kai-Min Hsieh: Experiment (supporting). Jeng-Chung Chen: Supervision (supporting). Anton Frisk Kockum: Supervision (supporting); Writing -- review \& editing (supporting). Guin-Dar Lin: Supervision (supporting); Writing -- review (supporting). Zhi-Rong Lin: Supervision (supporting); review (supporting). Ping-Yi Wen: Supervision (supporting); review (supporting). Io-Chun Hoi: Supervision (Lead); Writing -- review \& editing (Lead).

\section*{Data Availability Statement}

The data that support the findings of this study are available from the corresponding author upon reasonable request.


\section*{REFERENCES}
\bibliography{reference}

\begin{thebibliography}{16}
\bibitem [{Not()}]{Note}%
  \BibitemOpen
  \href@noop {} {}\bibinfo {note} {The nonradiative relaxation rate is absorbed into $\Gamma_{\phi}$.}\BibitemShut {Stop}%
\end{thebibliography}

\end{document}


\title[]{Supplementary Material for ``Tunable frequency conversion and comb generation with a superconducting artificial atom''}

\author{Fahad Aziz}
\homepage{These authors contributed equally to this work.}
\affiliation{Department of Physics, National Tsing Hua University, Hsinchu 30013, Taiwan.}
\affiliation{Department of Physics, City University of Hong Kong, Kowloon, Hong Kong SAR 999077, China}

\author{Zhengqi~Niu}
\homepage{These authors contributed equally to this work.}
\affiliation{State Key Laboratory of Materials for Integrated Circuits, Shanghai Institute of Microsystem and Information Technology (SIMIT), Chinese Academy of Science, Shanghai 200050, China}

\author{Tzu-Yen~Hsieh}
\homepage{These authors contributed equally to this work.}
\affiliation{Department of Physics and CQSE, National Taiwan University, Taipei 10617, Taiwan}%

\author{Kuan Ting~Lin}
\homepage{These authors contributed equally to this work.}
\affiliation{Trapped-Ion Quantum Computing Laboratory, Hon Hai Research Institute, Taipei 11492, Taiwan}

\author{Yu-Huan Huang}
\affiliation{Department of Physics, National Tsing Hua University, Hsinchu 30013, Taiwan.}

\author{Yen-Hsiang Lin}
\affiliation{Department of Physics, National Tsing Hua University, Hsinchu 30013, Taiwan.}
\affiliation{Center for Quantum Technology, National Tsing Hua University, Hsinchu 30013, Taiwan}

\author{Ching-Yeh Chen}
\affiliation{Department of Physics, National Tsing Hua University, Hsinchu 30013, Taiwan.}

\author{Yu-Ting Cheng}
\affiliation{Department of Physics, City University of Hong Kong, Kowloon, Hong Kong SAR 999077, China}

\author{Kai-Min Hsieh}
\affiliation{Department of Physics, City University of Hong Kong, Kowloon, Hong Kong SAR 999077, China}

\author{Jeng-Chung Chen}
\affiliation{Department of Physics, National Tsing Hua University, Hsinchu 30013, Taiwan.}
\affiliation{Center for Quantum Technology, National Tsing Hua University, Hsinchu 30013, Taiwan}

\author{Anton Frisk Kockum}
\affiliation{Department of Microtechnology and Nanoscience, Chalmers University of Technology,  412 96 Gothenburg, Sweden}

\author{Guin-Dar Lin}
\affiliation{Department of Physics and CQSE, National Taiwan University, Taipei 10617, Taiwan}
\affiliation{Trapped-Ion Quantum Computing Laboratory, Hon Hai Research Institute, Taipei 11492, Taiwan}
\affiliation{Physics Division, National Center for Theoretical Sciences, Taipei 10617, Taiwan}

\author{Zhi-Rong~Lin}
\homepage{Authors to whom correspondence should be addressed:\\
zrlin@mail.sim.ac.cn; \\
eighchin@ccu.edu.tw; \\
and iochoi@cityu.edu.hk}
\affiliation{State Key Laboratory of Materials for Integrated Circuits, Shanghai Institute of Microsystem and Information Technology (SIMIT), Chinese Academy of Science, Shanghai 200050, China}

\author{Ping-Yi Wen}
\homepage{Authors to whom correspondence should be addressed:\\
zrlin@mail.sim.ac.cn; \\
eighchin@ccu.edu.tw; \\
and iochoi@cityu.edu.hk}
\affiliation{Department of Physics, National Chung Cheng University, Chiayi 621301, Taiwan}

\author{Io-Chun Hoi}
\homepage{Authors to whom correspondence should be addressed:\\
zrlin@mail.sim.ac.cn; \\
eighchin@ccu.edu.tw; \\
and iochoi@cityu.edu.hk}
\affiliation{Department of Physics, National Tsing Hua University, Hsinchu 30013, Taiwan.}
\affiliation{Department of Physics, City University of Hong Kong, Kowloon, Hong Kong SAR 999077, China}

\maketitle


\tableofcontents

\renewcommand{\thefigure}{S\arabic{figure}}
\renewcommand{\thesection}{S\arabic{section}}
\renewcommand{\theequation}{S\arabic{equation}}
\renewcommand{\thetable}{S\arabic{table}}
\renewcommand{\bibnumfmt}[1]{[S#1]~}
\renewcommand{\citenumfont}[1]{S#1}


\newpage

\section{Detailed experimental setup}
\label{sec:1}

\begin{figure}
	\includegraphics[width=\linewidth]{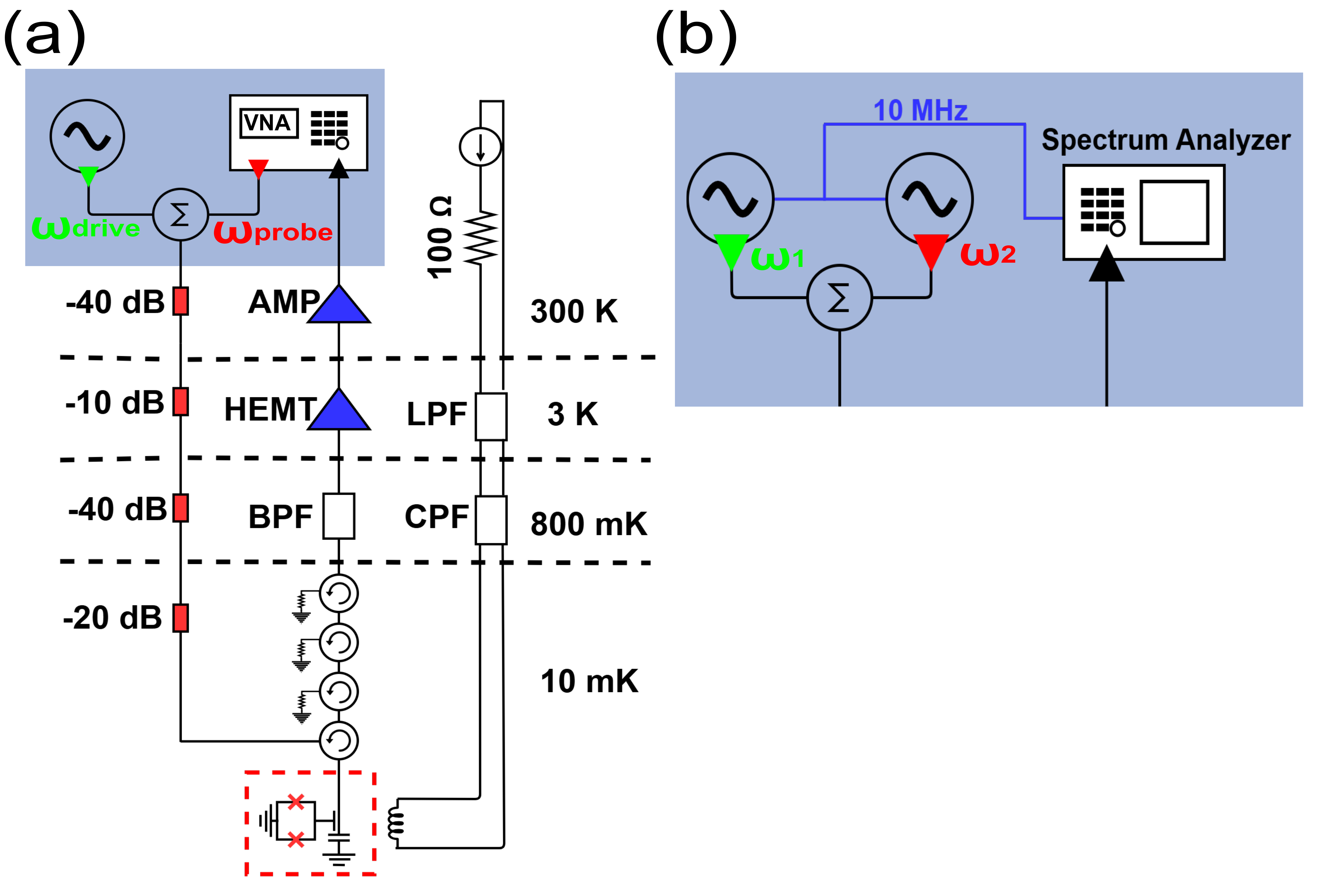}
	\caption{Schematic of the experimental setups for
	(a) measuring the reflection coefficient and 
	(b) the power spectral density. The setup outside the blue colored box in (b) is identical to that outside the blue colored box in (a). The red dashed box in (a) marks the transmon qubit.
	\label{fig:1}}
\end{figure}

The detailed experimental setup is presented in \figpanel{fig:1}{a}. To characterize the superconducting artificial atom, we measure the reflection coefficient using a vector network analyzer (VNA), with further details provided in the following section. A probe field ($\omega_{\rm probe}$), attenuated at multiple stages within the dilution refrigerator [indicated by solid red rectangles in \figpanel{fig:1}{a}], is directed through a cryogenic circulator and interacts with the artificial atom. The reflected output field is subsequently isolated and guided through a series of components, including cryogenic circulators, a band-pass filter (BPF), a high-electron-mobility transistor (HEMT), and a room-temperature amplifier (AMP). The amplified signal is finally recorded by the VNA. Additionally, a superconducting magnet is positioned at the centre of the sample box to enable tuning of the $\ket{0} \leftrightarrow \ket{1}$ transition frequency $\omega_{10}$~\cite{Hoi2015}. The magnetic field is controlled by adjusting the current supplied from a room-temperature direct-current (DC) source, followed by a low-pass filter (LPF) and a copper powder filter (CPF) to suppress noise.

For the two-tone measurement, we employ a similar approach, utilizing a single radio-frequency (RF) source generating a drive tone at frequency $\omega_{\rm drive} \equiv \omega_1$, combined with the setup via an RF combiner at room temperature, as also shown in \figpanel{fig:1}{a}. The rest of the experimental setup for measuring the reflection coefficient remains unchanged. This RF source is specifically used to drive the $\ket{0} \leftrightarrow \ket{1}$ transition (see \secref{sec:3} for details) on resonance, i.e., $\omega_{\rm drive} = \omega_{10}$, allowing us to characterize the anharmonicity and observe the well-known Mollow triplet~\cite{astafiev2010resonance, Wen2018}.

In \figpanel{fig:1}{b}, we show how the setup is changed from \figpanel{fig:1}{a} to measure the power spectral density; here, we utilize a spectrum analyzer. This configuration involves two radio-frequency (RF) sources, generating tones at frequencies $\omega_1$ and $\omega_2$, which are combined using an RF combiner. Both RF sources are synchronized to a reference signal provided by the spectrum analyzer. In \figref{fig:1}, green and red triangles represent the input signals, while the black triangles denote the reflected output signal, following the same path as described in \figpanel{fig:1}{b}. The output signal is subsequently measured using the power spectrum analyzer.


\section{Qubit characterization through single-tone spectroscopy}
\label{sec:2}

\begin{figure}
	\includegraphics[width=\linewidth]{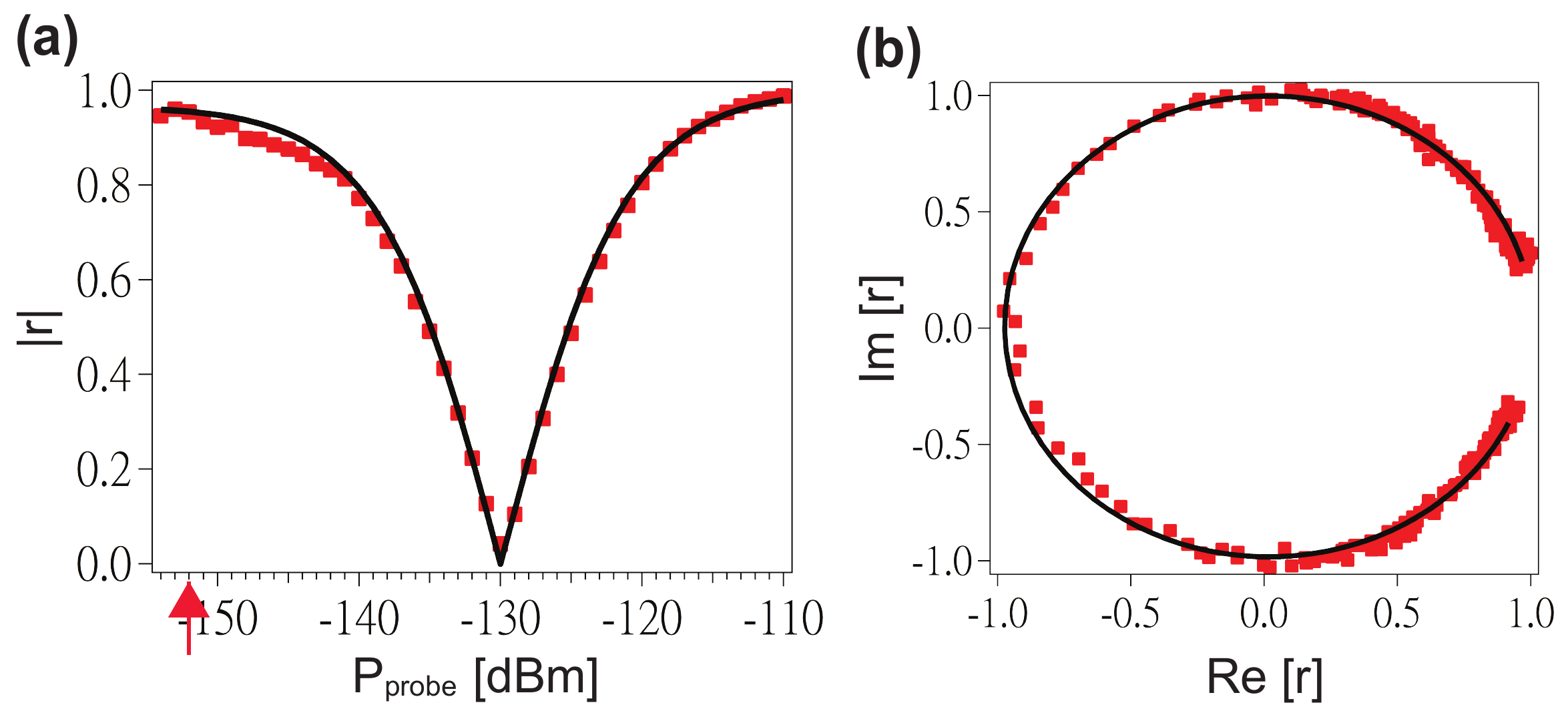}
	\caption{Single-tone spectroscopy. 
	(a) Magnitude of reflection coefficient, $|r|$, as a function of probe power at $\omega_{\rm probe} = \omega_{10}$. 
	(b) Reflection coefficient plotted in the IQ plane at $P_{\rm probe} = \unit[-152]{dBm}$. In these plots, red markers represent experimental data and solid black curves represent fits.
	\label{fig:2}}
\end{figure}

We performed single-tone spectroscopy to characterize the fundamental parameters of the qubit (the superconducting artificial atom). First, we biased the qubit at $\omega_{10} / 2 \pi = \unit[4.82]{GHz}$ using a DC source. Then we sent in a probe tone with different powers at this resonant frequency using a VNA and measured the reflected signal. In \figpanel{fig:2}{a}, we show the measured magnitude of the reflection coefficient, $|r|$, as a function of the probe power $P_{\rm probe}$. At $P_{\rm probe} = \unit[-130]{dBm}$, we observe $|r| = 0$. In the figure, red markers represent experimental data, while the solid black curve is the power fit using the method described in Ref.~\cite{Cheng2024}. The red arrow in \figpanel{fig:2}{a} indicates the power at which we characterize the fundamental parameters of the qubit, by performing a circle fit~\cite{Probst2015, lu2021characterizing} in the IQ plane as shown in \figpanel{fig:2}{b} (the solid black curve is the circle fit). All parameters extracted from these experiments are summarized in \tabref{tab:1}.

\begin{table*}
\centering
\begin{tabular}{| c | c | c | c | c | c | c  | c | c | c | c |}
\hline
$E_C / h$  & $E_J / h$ & $E_J / E_C$ & $\omega_{10}/ 2 \pi$  & $\omega_{21}/ 2 \pi$ &  $\Gamma_{10} / 2 \pi$  & $\Gamma_{\phi}/2\pi$ & $\gamma_{10} / 2 \pi$ & Gain & Attenuation \\
\hline
[MHz] & [GHz]  & - & [GHz] & [GHz]  & [MHz] & [MHz] & [MHz] & [dB] & [dB] \\
\hline
320 & 10.32 & 32.25 & 4.82 & 4.5 & 44.20  & 0.37 & 22.47 & 92.3 & -125 \\
\hline
\end{tabular}
\caption{Extracted fundamental parameters for the superconducting artificial atom in our experiment. We determined the transition frequency $\omega_{10}$, the relaxation rate $\Gamma_{10}$, the pure dephasing rate $\Gamma_{\phi}$~\cite{Note}, and the decoherence rate $\gamma_{10}$ by perfoming single-tone spectroscopy and using the circle-fit technique~\cite{Probst2015, lu2021characterizing}. We extracted the anharmonicity $\omega_{21} - \omega_{10}$, which is approximately equal to the charging energy $E_C$~\cite{Koch2007}, from two-tone spectroscopy experiments with results shown in \figpanel{fig:3}{a}. The gain and attenuation are extracted by performing a power fit on the data from single-tone spectroscopy following the method described in Ref.~\cite{Cheng2024}.
\label{tab:1}}
\end{table*}


\section{Two-tone spectroscopy and dressed states}
\label{sec:3}

\begin{figure}
	\includegraphics[width=\linewidth]{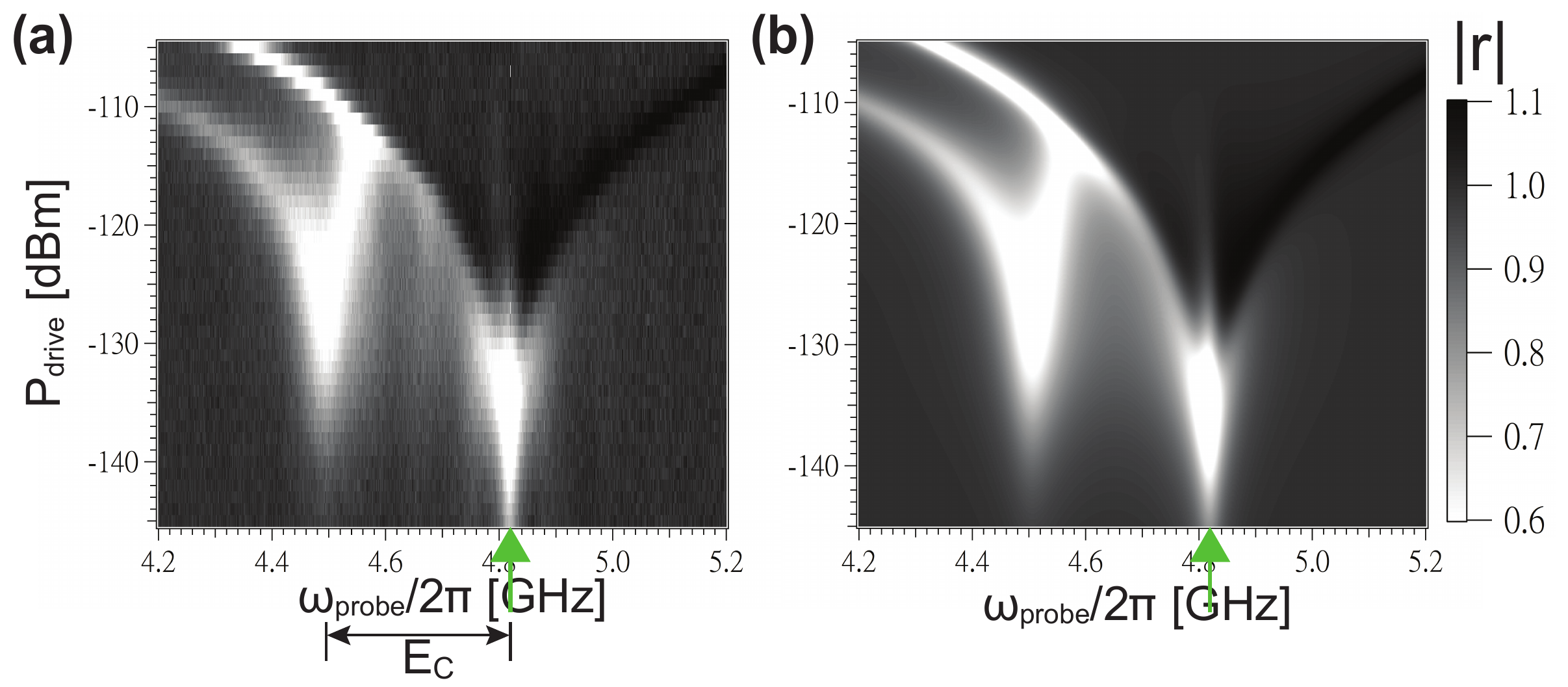}
	\caption{Two-tone spectroscopy. 
	(a) Measured reflection coefficient $|r|$ as a function of probe frequency $\omega_{\rm probe}$ and drive power $P_{\rm drive}$. The drive frequency is fixed at the qubit's resonance frequency $\omega_{10} / 2 \pi = \unit[4.82]{GHz}$ (marked by the green arrow), while the probe power is maintained at $P_{\rm probe} = \unit[-145]{dBm}$.
	(b) Theoretical simulation of the experimental results in (a), using the extracted parameters in \tabref{tab:1}.
	\label{fig:3}}
\end{figure}

Next, we measured the reflection coefficient $|r|$ in two-tone spectroscopy, yielding the results shown in \figpanel{fig:3}{a}.  We fixed the driving frequency of the RF source at $\omega_{\rm drive} = \omega_{10} / 2 \pi = \unit[4.82]{GHz}$, marked by the green arrow on the x axis, and varied the drive power on the y axis from $\unit[-145]{dBm}$ to $\unit[-105]{dBm}$. The x axis represents the probe frequency $\omega_{\rm probe}$; the probe signal was maintained at a fixed weak probe power of $P_{\rm probe} = \unit[-145]{dBm}$.

In \figpanel{fig:3}{a}, we observe the $\ket{0} \leftrightarrow \ket{1}$ and $\ket{1} \leftrightarrow \ket{2}$ transition frequencies at \unit[4.82]{GHz} and \unit[4.5]{GHz}, respectively, in the low-drive-power regime. The measured anharmonicity is thus $E_C / h \approx \unit[320]{MHz}$. As the drive power increases to $\unit[-130]{dBm}$, we observe the splitting of the $\ket{0} \leftrightarrow \ket{1}$ transition frequency into dressed states, generating the Mollow triplet. At a drive power of $\unit[-120]{dBm}$, we observe the Autler-Townes splitting (ATS) of the $\ket{1} \leftrightarrow \ket{2}$ transition frequency. The experimental results in \figpanel{fig:3}{a} agree well with the theoretical simulations in \figpanel{fig:3}{b}, which are based on the extracted parameters.


\section{Theoretical model}
\label{sec:4}

In this work, we consider a semi-infinite waveguide terminated by an anti-node mirror, with an $M$-level transmon ($M = 5$) placed at the end. This configuration enables strong transmon-waveguide coupling~\cite{Lin2019, aziz2025, Lin_2025}. We send in two coherent microwave tones, characterized by carrier frequencies $\omega_1$ and $\omega_2$ and Rabi frequencies $\Omega_1$ and $\Omega_2$, respectively, to interact with the transmon and induce a wave-mixing process. Note that the number of transmon levels $M$ must be chosen large enough to ensure that the population in the $(M+1)$th level remains negligible for any strength of the applied fields, thereby satisfying the cutoff condition. We chose $M = 5$ because, in all the experimental cases considered, the population of the sixth energy level was negligible. Thus, truncating the transmon to five levels offers an accurate and computationally efficient description of the system dynamics.


\subsection{Hamiltonian}

This system is described by the Hamiltonian
\begin{equation}
H = H_s + H_b + H_{\rm int} + H_{\rm mw_1} + H_{\rm mw_2}, 
\end{equation}
where the energies of the transmon and the waveguide photons are given by
\begin{equation}
H_s = \sum_{m = 1}^{M - 1} \hbar \omega_m \sigma_{mm}
\label{eq:H_s}
\end{equation}
and
\begin{equation}
H_b = \int_0^\infty d\omega \: \hbar \omega a_\omega^\dag a_\omega,
\label{eq:H_b}
\end{equation}
respectively. Here, $\hbar \omega_m$ is the energy of the $m$th excited state of the transmon (the energy of the ground state is set to zero), with the atomic projection operator $\sigma_{mm} = \ketbra{m}{m}$ for $m = 1, 2, \ldots, M-1$. The operator $a_\omega^\dag$ ($a_\omega$) represents the creation (annihilation) of a waveguide photon with frequency $\omega$, obeying the bosonic commutation relation $\mleft[ a_\omega , a_{\omega'}^\dag \mright] = \delta (\omega - \omega')$.

The transmon-waveguide interaction is described by the Hamiltonian
\begin{equation}
H_{\rm int} = i \hbar \sum_{m = 1}^{M - 1} \int_0^\infty d\omega \: \sqrt{m} g(\omega) \sigma_{m, m-1} a_\omega + \text{H.c.},
\label{eq:H_int}
\end{equation}
where $\sqrt{m} g(\omega)$ is the coupling strength between the $m$th transition of the transmon and a waveguide photon with frequency $\omega$~\cite{Koch2007}. The atomic ladder operator is expressed as $\sigma_{jk} = \ketbra{j}{k}$, and H.c.~denotes the Hermitian conjugate. Finally, the terms
\begin{equation}
H_{\rm mw_1} = \hbar \sum_{m = 1}^{M - 1} \frac{\sqrt{m} \Omega_1}{2} \mleft( \sigma_{m, m-1} e^{-i \omega_1 t} + \text{H.c.} \mright)
\label{eq:H_mw1}
\end{equation}
and
\begin{equation}
H_{\rm mw_2} = \hbar \sum_{m = 1}^{M - 1} \frac{\sqrt{m} \Omega_2}{2} \mleft( \sigma_{m, m-1} e^{-i \omega_2 t} + \text{H.c.} \mright),
\label{eq:H_mw2}
\end{equation}
describe the interaction of the microwave fields with the transmon~\cite{Kockum2013, Wen2018}.


\subsection{Master equation}

Using the standard approach to trace out the photonic modes of the waveguide~\cite{Lin2019, Lin_2025, Lehmberg1970}, we obtain the Born--Markov master equation in the frame rotating at frequency $\omega_s = (\omega_1 + \omega_2) / 2$:
\begin{align}
\frac{d\rho}{dt} & = i \sum_{m = 1}^{M - 1} \mleft( m \omega_s - \omega_m \mright) \mleft[ \sigma_{mm}, \rho \mright] \nonumber \\
& + i \sum_{m = 1}^{M - 1} \frac{\sqrt{m} \Omega_1}{2} \mleft[ \sigma_{m, m-1} e^{-i \delta t} + \sigma_{m-1, m} e^{i \delta t}, \rho \mright] \nonumber \\
& + i \sum_{m = 1}^{M - 1} \frac{\sqrt{m} \Omega_2}{2} \mleft[ \sigma_{m, m-1} e^{i \delta t} + \sigma_{m-1, m} e^{-i \delta t}, \rho \mright] \nonumber \\
& + \sum_{m = 1}^{M - 1} \sum_{n = 1}^{M - 1} \frac{\sqrt{mn} \Gamma_{m, m-1}}{2} \mleft( \mleft[ \sigma_{m-1, m} \rho, \sigma_{n, n-1} \mright] + \mleft[ \sigma_{n-1, n}, \rho \sigma_{m, m-1} \mright] \mright) \nonumber \\
& + \sum_{m = 1}^{M - 1} \sum_{n = 1}^{M - 1} \frac{mn \Gamma_\phi}{2} \mleft( \mleft[ \sigma_{mm} \rho, \sigma_{nn} \mright] + \mleft[ \sigma_{nn}, \rho \sigma_{mm} \mright] \mright) . 
\label{master equation}
\end{align}
Here, $\delta = (\omega_1 - \omega_2) / 2$ is half the detuning between the two microwave fields, and the atomic relaxation rate is denoted by $\Gamma_{m, m-1} = 2\pi g^2 (\omega_m - \omega_{m - 1})$. The last term in the master equation is added manually to account for the effect of pure dephasing, described by the dephasing rate $m^2 \Gamma_{\phi}$ of the $m$th excited state. The driving terms, i.e., the second and third lines in the master equation, introduce the oscillating factors $e^{\pm i \delta t}$. These oscillating factors result from the interference between the two microwave fields of distinct frequencies, producing a beat frequency that periodically modulates the transmon dynamics.

To address this periodicity and solve~\eqref{master equation}, we expand each atomic component using a Fourier series:
\begin{equation}
\expec{\sigma_{mn}(t)} = \sum_{l = - \infty}^\infty \expec{\sigma_{mn}(t)}^{(l)} e^{i l \delta t}.
\label{Fourier decomposition of sigma}
\end{equation}
By leveraging the beat frequency and the Fourier expansion, we isolate the contributions of different harmonic orders; this leads to the optical Bloch equations for the Fourier amplitudes of order $l$~\cite{Ficek1993}:
\begin{align}
\frac{d \expec{\sigma_{mn}(t)}^{(l)}}{dt} & = \mleft[ i \mleft( \Theta_{mn} - l \delta \mright) - \Xi_{mn} \mright] \expec{\sigma_{mn}(t)}^{(l)} \nonumber \\
& + i \frac{\sqrt{n}}{2} \mleft( \Omega_1 \expec{\sigma_{m, n-1}(t)}^{(l-1)} + \Omega_2 \expec{\sigma_{m, n-1}(t)}^{(l+1)} \mright) \nonumber \\
& + i \frac{\sqrt{n+1}}{2} \mleft( \Omega_1 \expec{\sigma_{m, n+1}(t)}^{(l+1)} + \Omega_2 \expec{\sigma_{m, n+1}(t)}^{(l-1)} \mright) \nonumber \\
& - i \frac{\sqrt{m+1}}{2} \mleft( \Omega_1 \expec{\sigma_{m+1, n}(t)}^{(l-1)} + \Omega_2 \expec{\sigma_{m+1, n}(t)}^{(l+1)} \mright) \nonumber \\
& - i \frac{\sqrt{m}}{2} \mleft( \Omega_1 \expec{\sigma_{m-1, n}(t)}^{(l+1)} + \Omega_2 \expec{\sigma_{m-1, n}(t)}^{(l-1)} \mright) \nonumber \\
& + \frac{\sqrt{(m+1) (n+1)}}{2} \mleft( \Gamma_{n+1, n} + \Gamma_{m+1, m} \mright) \expec{\sigma_{m+1, n+1}(t)}^{(l)} .
\label{Bloch-equations in Fourier form}
\end{align}
The effective frequencies and decoherence rates here are given by
\begin{equation}
\Theta_{mn} = (n - m) \omega_s - (\omega_n - \omega_m)
\label{eq:effective freqency}
\end{equation}
and
\begin{equation}
\Xi_{mn} = \frac{\mleft( n \Gamma_{n, n-1} + m \Gamma_{m, m-1} \mright)}{2} + \mleft( m n - \frac{1}{2} m^2 - \frac{1}{2} n^2 \mright)  \Gamma_{\phi} ,
\label{eq:effective decoherence rate}
\end{equation}
respectively.


\subsection{Coherent and incoherent spectra}

To analyze the wave-mixing process, we measure the fluorescence spectrum using a spectrum analyzer. The spectrum, $S(\omega)$, can be determined using the expression
\begin{equation}
S(\omega) = \frac{1}{2\pi} \lim_{t \rightarrow \infty} \text{Re} \int_{-\infty}^\infty d\tau \: \frac{\expec{V_{\rm sc}^\dag (t + \tau) V_{\rm sc} (t)}}{2 Z_0} e^{- i \omega \tau},
\label{eq:definition of S(omega)}
\end{equation}
where $Z_0$ denotes the characteristic impedance of the transmission line, and the scattered signal is given by
\begin{equation}
V_{\rm sc} (t) = V_{\rm out} (t) - V_{\rm in} (t) = - i \sqrt{\frac{\hbar Z_0}{4\pi}} \int_0^\infty d\omega \: \sqrt{\omega} \tilde{a}_\omega (t) e^{-i (\omega - \omega_s) t}.
\label{V_SC}
\end{equation}
Here, $V_{\rm in}$ and $V_{\rm out}$ represent the input and output signal~\cite{Lin2019, Peropadre2013}, and $\tilde{a}_\omega (t) = a(t) e^{i (\omega - \omega_s) t}$ is the slowly varying photonic operator with $\dot{\tilde{a}}_\omega (t) \approx 0$. After solving the Heisenberg equation for $\tilde{a}_\omega (t)$, we find
\begin{equation}
\tilde{a}_{\omega} (t) = - \sum_m \sqrt{m} g(\omega) \int_0^t dt' \tilde{\sigma}_{m-1, m} (t') e^{i \mleft[ \omega - \mleft( \omega_m - \omega_{m-1} \mright) + \omega_s \mright] t'},
\label{a_omega}
\end{equation}
where $\tilde{\sigma}_{m-1, m} (t) = \sigma_{m-1, m} (t) e^{i \mleft[ \mleft(\omega_m - \omega_{m-1} \mright) - \omega_s \mright] t}$. By substituting \eqref{a_omega} into \eqref{V_SC}, the fluorescence spectrum is obtained as
\begin{equation}
S(\omega) = \frac{\hbar}{4\pi} \sum_{m, n} C_{mn} \lim_{t \rightarrow \infty} \text{Re} \int_0^\infty d\tau \: \expec{\tilde{\sigma}_{m, m-1} (t + \tau ) \tilde{\sigma}_{n-1, n} (t)} e^{-i \mleft( \omega - \omega_s \mright) \tau} ,
\label{S(omega)}
\end{equation}
with the coefficients $C_{mn} = \sqrt{mn \mleft( \omega_m - \omega_{m-1} \mright) \mleft( \omega_n - \omega_{n-1} \mright) \Gamma_{m, m-1} \Gamma_{n, n-1}}$.

By introducing the fluctuation operators
\begin{equation}
\Delta \tilde{\sigma}_{m, n} (t) = \tilde{\sigma}_{m, n} (t) - \expec{\tilde{\sigma}_{m, n} (t)} ,
\end{equation}
we can decompose the two-time correlation function in \eqref{S(omega)} as
\begin{equation}
\expec{\tilde{\sigma}_{m, m-1} (t + \tau) \tilde{\sigma}_{n-1, n} (t)} = \expec{\Delta \tilde{\sigma}_{m, m-1} (t + \tau) \Delta \tilde{\sigma}_{n-1, n} (t)} - \expec{\tilde{\sigma}_{m, m-1} (t + \tau)} \expec{\tilde{\sigma}_{n-1, n} (t)} . 
\end{equation}
This decomposition leads to the incoherent contribution
\begin{equation}
S_{\rm inco}(\omega) = \frac{\hbar}{4 \pi} \sum_{m, n} C_{mn} \lim_{t \rightarrow \infty} \text{Re} \int_0^\infty d\tau \: \expec{\Delta \tilde{\sigma}_{m, m-1} (t + \tau) \Delta \tilde{\sigma}_{n-1, n} (t)} e^{-i \mleft( \omega - \omega_s \mright) \tau}
\label{S_inco}
\end{equation}
and the coherent contribution
\begin{equation}
S_{\rm co}(\omega) = \frac{\hbar}{4 \pi} \sum_{m, n} C_{mn} \lim_{t \rightarrow \infty} \text{Re} \int_0^\infty d\tau \: \expec{\tilde{\sigma}_{m, m-1} (t + \tau)} \expec{\tilde{\sigma}_{n-1, n} (t)} e^{-i \mleft( \omega - \omega_s \mright) \tau}.
\label{S_co}
\end{equation}

The fluorescence spectrum is thus divided into incoherent and coherent components. For the incoherent spectrum, \eqref{S_inco}, the two-time correlation function can be calculated by using the quantum regression theorem~\cite{Ficek1993, Scully1997} and by solving the optical Bloch equations in \eqref{Bloch-equations in Fourier form}. For the coherent spectrum, we substitute the Fourier decomposition, \eqref{Fourier decomposition of sigma}, into \eqref{S_co} and obtain
\begin{equation}
S_{\rm co}(\omega) = \frac{\hbar}{4 \pi} \sum_{m, n} C_{mn} \sum_{l = -\infty}^\infty \expec{\tilde{\sigma}_{m, m-1}}_{\rm ss}^{(l)} \expec{\tilde{\sigma}_{n-1, n}}_{\rm ss}^{(-l)} \pi \delta \mleft[ \omega - \mleft( \omega_s + l \delta \mright) \mright] ,
\label{S_co-1}
\end{equation}
where $\expec{\tilde{\sigma}_{m, n}}_{\rm ss}^{(l)}$ represents the $l$th-order Fourier component of the periodic steady-state solution $\expec{\tilde{\sigma}_{m, n} (t)}$ with oscillation frequency $l \delta$. 

To account for the fact that the delta function $\delta \mleft[ \omega - \mleft( \omega_s + l \delta \mright) \mright]$ corresponds to the idealized coherent spectrum arising from elastic photon scattering, but is not experimentally realizable due to finite measurement resolution, we introduce a decaying parameter $\epsilon_l$ into the exponential in \eqref{S_co-1}. This broadens the delta function into a Lorentzian function with width $\epsilon_l$, allowing us to describe the coherent spectrum as
\begin{equation}
S_{\rm co}(\omega) = \frac{\hbar}{4 \pi} \sum_{m, n} C_{mn} \sum_{l = - \infty}^\infty \expec{\tilde{\sigma}_{m, m-1}}_{\rm ss}^{(l)} \expec{\tilde{\sigma}_{n-1, n}}_{ss}^{(-l)} \frac{\epsilon_l}{\mleft[ \omega - \mleft( \omega_s + l \delta \mright) \mright]^2 + \epsilon_l^2}.
\label{S_co-2}
\end{equation}
Thus, the complete fluorescence spectrum can be expressed as $S(\omega) = S_{\rm inco}(\omega) + S_{\rm co}(\omega)$, representing the total energy distribution.

In our experiment, measurements were performed in runs with the pump on and with the pump off. To connect the theoretical predictions and experimental results, we start with the conversion formula from decibels (dB) to power in milliwatt (mW): $P = (1mW)10^{\frac{dB}{10}}$. Then, the power ratio between the pump-on and pump-off cases (powers $P_{\rm on}$ and $P_{\rm off}$, respectively) is
\begin{equation}
\frac{P_{\rm on}}{P_{\rm off}} = 10^{\frac{x}{10}} ,
\label{power ratio}
\end{equation}
where $x$ is the experimental raw data in units of dB. The left-hand side of \eqref{power ratio} can be written as
\begin{equation}
\frac{P_{\rm on}}{P_{\rm off}} = 1 + \frac{P_{\rm signal}}{P_{\rm off}} . 
\end{equation}

The fitting function used in this experiment is
\begin{equation}
F(\omega) = 1 + \mleft[ S_{\rm inco}(\omega) + S_{\rm co}(\omega) \mright] \frac{\Delta \omega_{\rm RBW}}{P_{\rm off}} ,
\label{fitting function}
\end{equation}
where $\Delta \omega_{\rm RBW}$ is the resolution bandwidth, which equals \unit[910]{kHz}, and $P_{\rm off}$ is the background noise power, which is treated as a fitting parameter. The normalized power spectral density (PSD$_n$) is calculated in decibels (dB) as
\begin{equation}
\text{PSD}_n = 10 \log F(\omega) .
\label{eq:psd_n}
\end{equation}
The fitting parameters for PSD$_n$ include $P_{\rm off}$, $\epsilon_i$ (width of the peak, where $i$ denotes the peak index), $k_1$, and $k_2$ (the scaling factors connecting input power and Rabi frequency of the driving fields).


\section{Frequency up- and down-conversion}
\label{sec:5}

\begin{figure}
	\includegraphics[width=\linewidth]{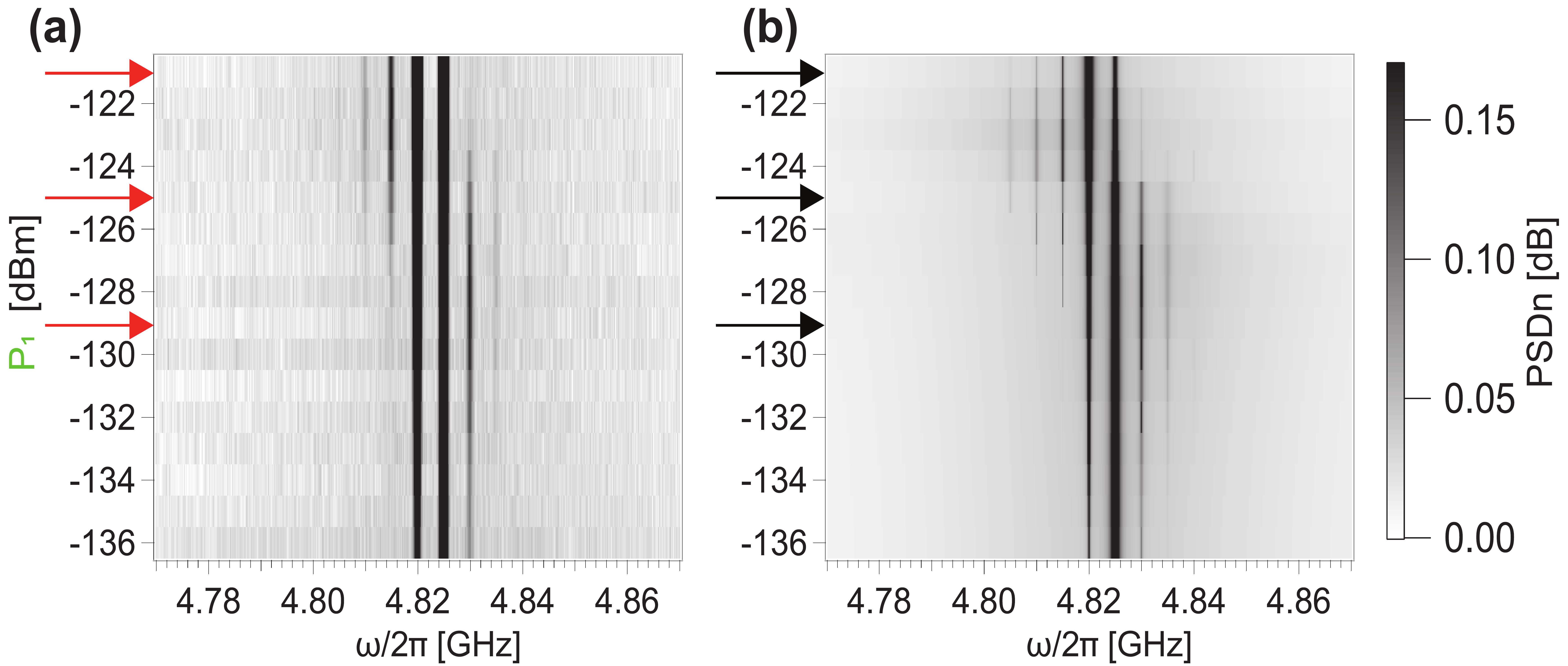}
	\caption{Normalized power spectral density PSD$_n$ as a function of spectrum analyzer frequency $\omega / 2\pi$ and RF-source power $P_1$, with input frequencies fixed at $\omega_1 / 2\pi = \unit[4.82]{GHz}$ and $\omega_2 / 2\pi = \unit[4.825]{GHz}$, and power $P_2$ fixed at $\unit[-125]{dBm}$.
	(a) Experimental results showing the generation of several new frequency peaks, beyond those at $\omega_1$ and $\omega_2$, when tuning $P_1$. Three power regimes are identified, illustrating the control of frequency conversion through $P_1$.
	(b) The theoretical results, obtained as described in \secref{sec:4}, align with the experimental observations, confirming that frequency conversion can be tuned by adjusting $P_1$.
	\label{fig:4}}
\end{figure}

In \figref{fig:4}, we show the normalized power spectral density PSD$_n$ as a function of spectrum analyzer frequency $\omega / 2\pi$ and the input power $P_1$. We varied $P_1$ from $\unit[-136]{dBm}$  to $\unit[-121]{dBm}$ while maintaining a fixed input frequency of $\omega_1 / 2\pi = \unit[4.82]{GHz}$ and a fixed input power of $P_2 = \unit[-125]{dBm}$ at a constant frequency of $\omega_2 / 2\pi = \unit[4.825]{GHz}$. The three horizontal colored arrows indicate line cuts shown in Fig.~2 of the main text.

In \figpanel{fig:4}{a}, for low $P_1$ ($< \unit[-125]{dBm}$), we observe the generation of new frequency peaks at $2 \omega_2 - \omega_1$ and $3 \omega_2 - 2 \omega_1$. When $P_1 \approx P_2$, we observe the generation of new frequency peaks at $2 \omega_2 - \omega_1$, $3 \omega_2 - 2 \omega_1$, $2 \omega_1 - \omega_2$, and $3 \omega_1 - 2 \omega_2$, symmetrically. As $P_1$ increases ($> \unit[-125]{dBm}$), we observe new frequency peaks at $2 \omega_1 - \omega_2$ and $3 \omega_1 - 2 \omega_2$. These three power regimes, with their line cuts in the main text, demonstrate our control of frequency conversion to up-convert or down-convert the applied frequency by tuning the $P_1$ driving power. Figure~\figpanelNoPrefix{fig:4}{b} presents the corresponding theoretical results, which agree well with the experimental data.


\section{Generation of frequency combs}
\label{sec:7}


\subsection{Theoretical simulation on sweeping driving frequencies}

\begin{figure}
	\includegraphics[width=\linewidth]{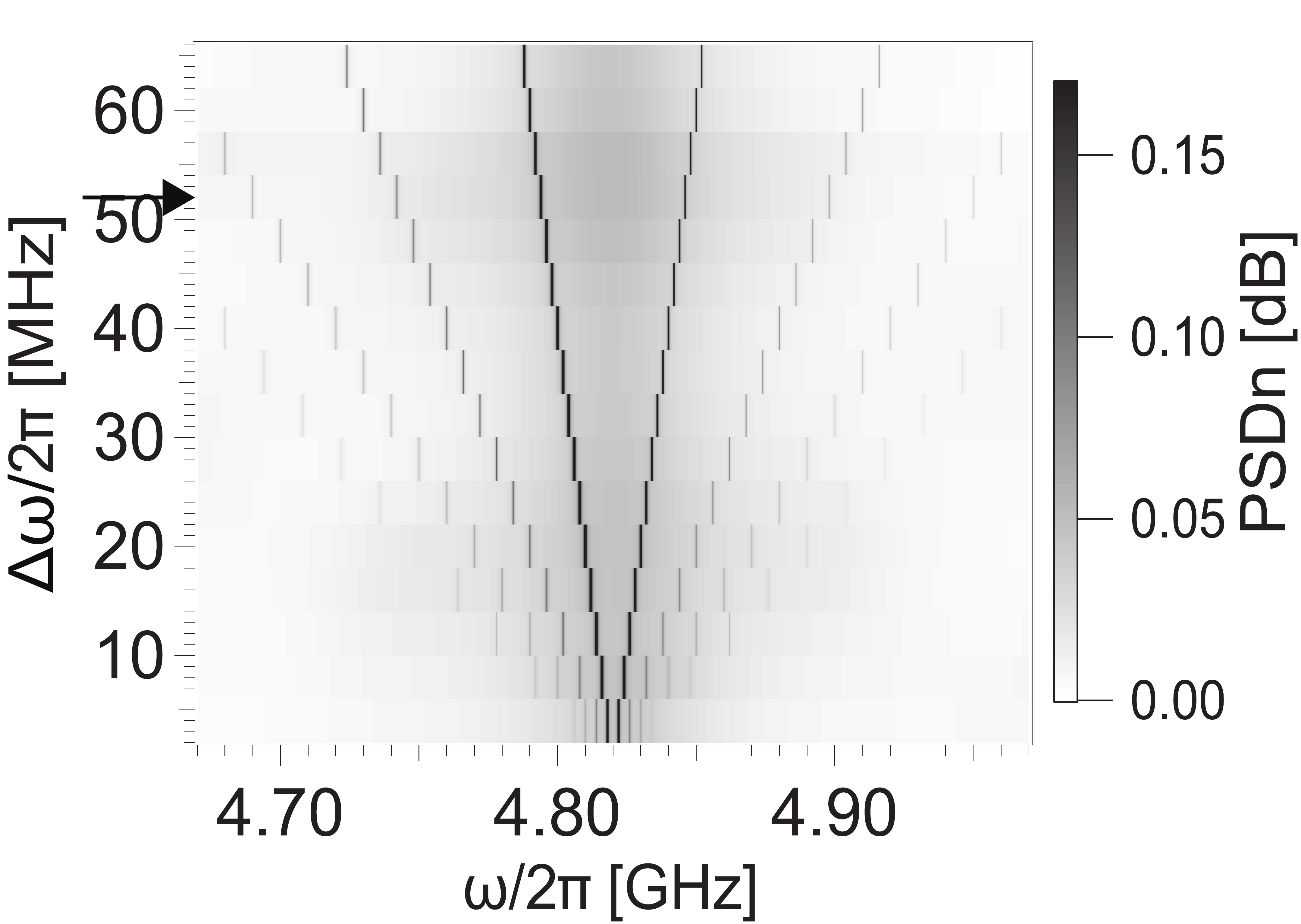}
	\caption{Normalized power spectral density PSD$_n$ as a function of the spectrum analyzer frequency $\omega / 2\pi$ and $\Delta \omega/2\pi$, the detuning  between the input frequencies $\omega_1$ and $\omega_2$. The input powers are fixed at $P_1 = P_2 = \unit[-123]{dBm}$, with the qubit biased at its resonant frequency $\omega_{10} / 2\pi = \unit[4.82]{GHz}$.
	\label{fig:5}}
\end{figure}

In \figref{fig:5}, we show the theoretical results corresponding to the experimental data in Fig.~4(a) of the main text. The theoretical prediction shows symmetric side peaks around the input frequencies, with peak positions depending on the detuning. These results illustrate the formation of a frequency comb under varying detuning conditions. The power spectral density exhibits six distinct peaks, symmetrically distributed around the input frequencies $\omega_1$ and $\omega_2$, consistent with the frequency-comb formation.

\subsection{Sweeping driving powers}

\begin{figure}
	\includegraphics[width=\linewidth]{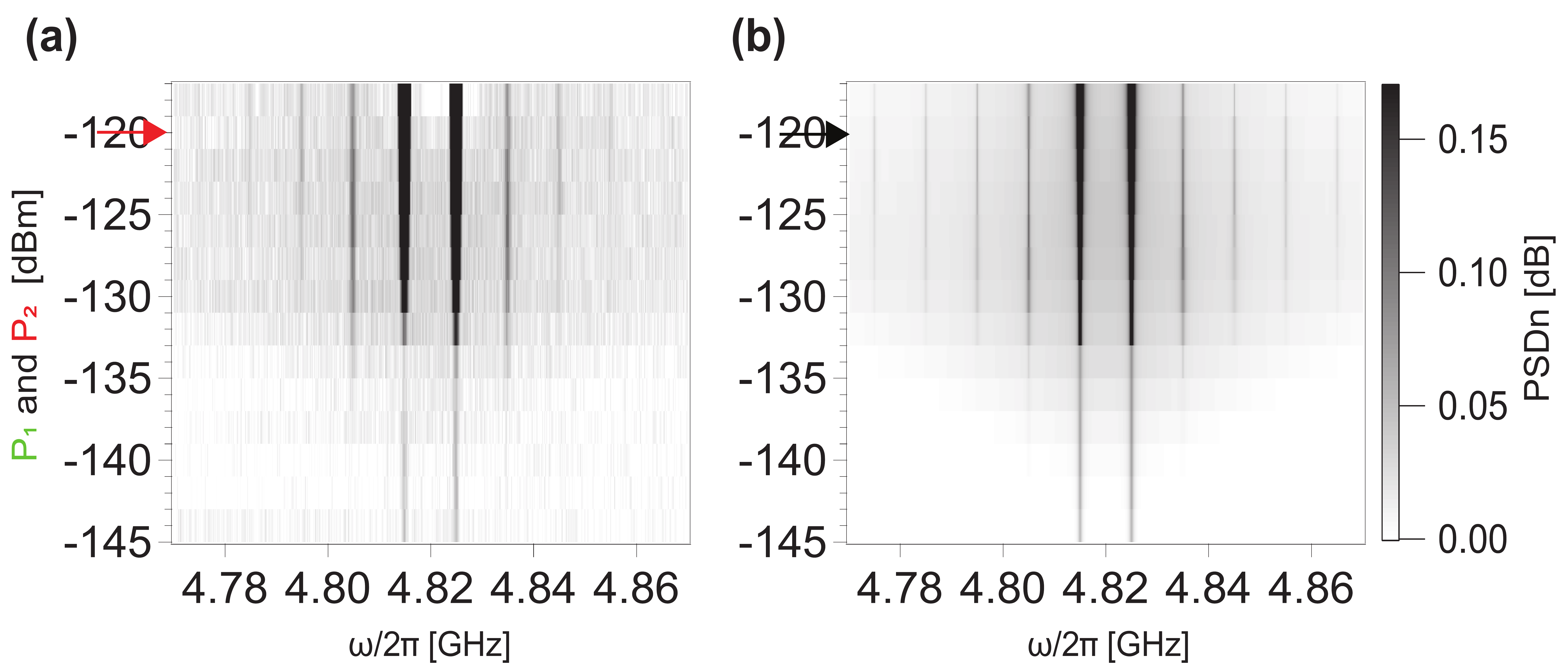}
	\caption{Normalized power spectral density PSD$_n$ as a function of the spectrum analyzer frequency and the input powers $P_1$ and $P_2$, with both frequencies fixed at $\omega_1 / 2\pi = \unit[4.815]{GHz}$ and $\omega_2 / 2\pi = \unit[4.825]{GHz}$ ($\Delta \omega / 2\pi = \unit[10]{MHz}$).
	Both (a) the experimental data and (b) the theoretical prediction show the enhancement of side peaks as the input powers increase from $\unit[-145]{dBm}$ to $\unit[-118]{dBm}$. At lower input powers, side peaks are too weak to be experimentally detected, while higher input powers lead to more pronounced and symmetric side peaks around the applied frequencies.
	\label{fig:6}}
\end{figure}

In \figref{fig:6}, we show the normalized power spectral density PSD$_n$ as a function of frequency and the input powers of $P_1$ and $P_2$. We simultaneously sweep both powers ($P_1$ and $P_2$) from $\unit[-145]{dBm}$ to $\unit[-114]{dBm}$. In this experiment, we fixed the drive frequencies at $\omega_1 / 2\pi = \unit[4.815]{GHz}$ and $\omega_2 / 2\pi = \unit[4.825]{GHz}$, resulting in a detuning of $\Delta \omega / 2\pi = \unit[10]{MHz}$. When the input powers are below $\unit[-135]{dBm}$, we observe only two peaks, corresponding to the applied frequencies. As we increase the power, we detect the generation of new peaks at equal detuning, symmetrical to the applied frequencies. The red and black arrows indicate the line cut discussed in Fig.~4(c) in the main text. Figure~\figpanelNoPrefix{fig:6}{a} presents the experimental results, while \figpanel{fig:6}{b} displays the theoretical results from simulations based on \secref{sec:4}; the two show good agreement.

\bibliography{Ref}